# Algorithms and Data Structures to Accelerate Network Analysis

Jordi Ros-Giralt, Alan Commike, Peter Cullen, Richard Lethin
{giralt, commike, cullen, lethin}@reservoir.com
Reservoir Labs
632 Broadway Suite 803, New York, NY 10012

*Abstract -* As the sheer amount of computer generated data continues to grow exponentially, new bottlenecks are unveiled that require rethinking our traditional software and hardware architectures. In this paper we present five algorithms and data structures (long queue emulation, lockless bimodal queues, tail early dropping, LFN tables, and multiresolution priority queues) designed to optimize the process of analyzing network traffic. We integrated these optimizations on R-Scope, a high performance network appliance that runs the Bro network analyzer, and present benchmarks showcasing performance speed ups of 5X at traffic rates of 10 Gbps.

## I. Introduction

System wide optimization of network components like routers, firewalls, or network analyzers is complex as it involves the proper orchestration of at least hundreds of different algorithms and data structures interrelated in subtle ways. In these highly dynamic systems, bottlenecks quickly shift from one component to another forming a network of *micro-bottlenecks*. This makes it challenging to understand which elements should be further optimized to get that extra unit of performance. Moreover, these shifting micro-bottlenecks are interconnected in peculiar ways so that optimizing one of them can often lead to an overall degradation of performance. This is due to internal system nonlinearities such as those found in hierarchical memory architectures. For instance, while optimizing the transfer of packets from the wire to the application is known to be critical, in the limit pushing too many packets to the application is detrimental as packets that eventually need to be dropped will cause a net negative effect by thrashing the processors' local caches, increasing the overall cache miss ratios and hence decreasing system wide performance. The process of performance optimization should therefore be a meticulous one which requires making small but safe steps avoiding the pitfall of pursuing short term gains that can lead to a new and bigger bottleneck down the path.

In this paper we present five of such safe steps that have helped to optimize the performance of R-Scope, a high performance appliance that runs the network analyzer Bro at its core [1]. Each of these steps introduces a new algorithm or data structure designed to accelerate system wide performance, each one addressing a different shifting micro-bottleneck. While we use Bro to demonstrate the efficacy of these optimizations, they are of general purpose and so we believe these techniques can be generally applied to the problem of accelerating network analysis or, to some degree, to optimize other more active network components such as firewalls or routers.

## II. Algorithms and Data Structures

### A. Long Queue Emulation for Packet Forwarding

High performance network interface cards (NICs) help accelerate the process of moving packets from the wire to the application by using techniques such as *receive side scaling* (RSS), *zero copy*, *packet coalescence* or *kernel bypass*, among others [2]. These cards achieve higher performance by leveraging hardware at the cost of losing some degree of flexibility and programmability. For instance, one common element of rigidity found in HPC NICs is the amount of memory embedded in their chip, which limits the size of the rings used to temporarily hold packets as they are transferred to the application. As a result, temporary high bursts of traffic that cannot be handled fast enough by the application may overflow these hardware rings leading to packet drops.

A traditional way to address packet drops originated from a limited size ring (LSR) is to dedicate one or more dispatcher threads (DT) to move packets out of the ring and put them into one or more software queues connected to the application threads (AT) residing on the host. Because the host does not have the embedded memory restrictions of the NIC, the software queues effectively have unlimited size. Consequently, packet drops due to bursty traffic are eliminated provided that the dispatcher threads can move packets from the limited size rings (LSR) to the unlimited size queues (USQ) fast enough. This solution is illustrated in Fig. 1.

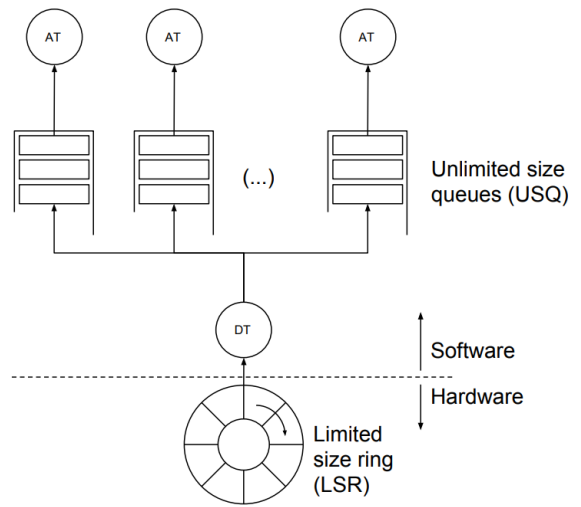

Fig. 1. Description of the dispatcher model.

---
Some of the data structures and algorithms in this paper have been patented.

While this solution seems sound at a high level, in the context of HPC the dispatcher thread introduces the following subtle but important performance penalties:

- *Packet read cache penalty.* If the dispatcher thread (DT) needs to read the packet—for instance, if it needs to compute the hash of the packet's IP tuple to decide which destination queue the packet should be forwarded to—then the packet will need to be loaded into the local cache. Since the application thread (AT) will also need to read the packet for its own processing, the dispatcher model requires loading a packet to the cache twice (one time on the DT's local cache and a second time on the AT's local cache). This negatively impacts performance because cache misses—which require accessing memory to fetch data—are typically ten times slower than cache hits. As a general principle, a good design should aim for a single cache load throughout the lifetime of each packet.
- *Descriptor read cache penalty.* Even if the DT does not need to read the packet—e.g., some implementations can extract the hash of the packet's IP tuple from the packet context information provided by the hardware—the DT will still need to load the packet descriptor onto its local cache. (A packet descriptor is a small software data structure part of all NIC drivers containing a pointer to the packet buffer and additional control metadata such as the the packet length or the hash of its IP tuple, among others.) In this case, during the lifetime of a packet, its descriptor needs to be loaded twice, once at the DT's local cache and a second time at the AT's local cache. Just like before, a good design should target one single cache load for each individual packet descriptor.
- *Memory and compute overhead.* Yet another overhead introduced by this approach is the additional memory and compute resources required to run the dispatcher threads themselves.

To avoid the above performance penalties, we propose to use *long queue emulation* (LQE), a simple but efficient technique that eliminates the overhead introduced by the dispatcher thread with the potential to also reduce packet drops.

The main concept behind LQE is to emulate the behavior of the dispatcher thread solution by folding the actions performed by the DT thread into the AT thread. Consider first the pseudocode of the DT and AT threads separately as implemented by the dispatcher model:

```
DtThread()
1    while true:
2        get alls packets from the front of LSR;
3        put the packets to the tail of USQ;

AtThread()
4    while true:
5        get one packet from the front of USQ;
6        process the packet;
```

While in the dispatcher thread solution the `DtThread()` and the `AtThread()` procedures are run on two independent threads, in the long queue emulation model we fold `DtThread()` into `AtThread()` as a single thread running the procedure `AtLqeThread()`:

```
AtLqeThread()
1    while true:
2        get all packets from the front of LSR;
3        put the packets to the tail of USQ;
4        get one packet from the front of USQ;
5        process the packet;
```

The key characteristic of the `AtLqeThread()` procedure is that it ensures all packets from the LSR ring are moved to the USQ queue before the next packet is processed, effectively giving the highest priority to the ring. This approach emulates the behavior of the dispatcher model with one single thread performing both the DT and the AT procedures. As a result, both packets and packet descriptors are loaded into the cache only once (at the AT's local cache) and there is no additional memory and compute overhead to maintain the dispatcher threads. We illustrate the LQE model in Fig. 2.

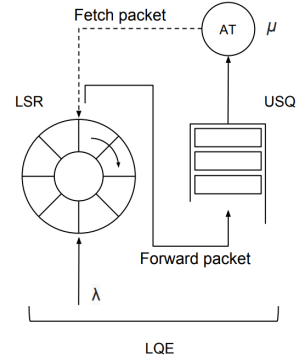

Fig. 2. Description of the long queue emulation model.

More formally, we describe the performance properties of the long queue emulation model in the following lemma:

*Lemma 1. Long queue emulation performance.* Let $\lambda$ and $\lambda_{max}$ be the average and the maximum packet arrival rate measured at the LSR ring, respectively. Assume for the sake of simplicity and without loss of generality, that the time to process a packet is constant, and let $\mu_{dt}$ and $\mu_{lqe}$ be the packet processing rate of the DT model and the LQE model, respectively—that is, $\mu_{dt}$ and $\mu_{lqe}$ correspond to one divided by the time it takes to execute line 6 in `AtThread()` and line 5 in the `AtLqeThread()`. If $s_{lsr}$ is the maximum number of packets that can be held in the LSR ring, then the following is true:

(1) $\mu_{lqe} > \mu_{dt}$.
(2) If $s_{lsr}/\lambda_{max} \geq 1/\mu_{lqe}$, the performance of the LQE model is superior to the performance of the DT model.
(3) If $s_{lsr}/\lambda_{max} < 1/\mu_{lqe}$ and $\lambda \geq \mu_{lqe}$, the performance of the LQE model is superior to the performance of the DT model.
(4) If $s_{lsr}/\lambda_{max} < 1/\mu_{lqe}$ and $\lambda < \mu_{lqe}$, the performance of the DT model is superior to the performance of the LQE model.

*Proof.* It's easy to see that $\mu_{lqe} > \mu_{dt}$ because the LQE model does not suffer from DT's performance penalties due to extra cache misses or the computational and memory overheads previously described.

To see that (2) is also true, notice that the computational cost of the procedures `AtLqeThread()` and `DtThread()` are the same except for the cost of processing a packet (assuming the cost of putting and getting packets from the ring and the queue is negligible compared to the cost of processing the packet). Since in the LQE model the time it takes to process one single packet is $1/\mu_{lqe}$, the maximum number of packets that can be inserted in the ring while the application thread is processing a packet is $\lambda_{max}/\mu_{lqe}$. Since $s_{lsr} \geq \lambda_{max}/\mu_{lqe}$, no packets are dropped and so both the long queue emulation and the dispatcher models deliver a packet drop probability equal to zero. Since from (1) we know that $\mu_{lqe} > \mu_{dt}$, we conclude that the LQE model uses less memory and compute resources and delivers the same packet drop probability as the DT model, making it the superior design.

If $s_{lsr} < \lambda_{max}/\mu_{lqe}$ and $\lambda \geq \mu_{lqe}$, then from queuing theory [3] we know that this leads to a permanently unstable regime where the USQ queue will grow indefinitely long and the system will not have a stationary distribution for either model. As a result, packets will be dropped at a rate $\lambda - \mu_{dt}$ and $\lambda - \mu_{lqe}$ for the DT and LQE model, respectively. From (1) we have $\mu_{lqe} > \mu_{dt}$, which makes also the LQE model a better design than the DT model.

Finally, if $s_{lsr} < \lambda_{max}/\mu_{lqe}$ and $\lambda < \mu_{lqe}$, then applying queuing theory again we know the system will be stable with traffic bursts leading to temporary increases of the queue size that the application thread will only be able to process if the LSR ring can accommodate for the burst. In the LQE model, since $s_{lsr} < \lambda_{max}/\mu_{lqe}$, the maximum burst will overflow the LSR ring, leading to packet drops at a rate $\lambda_{max} - \mu_{lqe}$ for the duration of the burst. In the DT model, however, such packet drops are eliminated as the dedicated DT thread can move the packets from the LSR ring to the USQ queue without drops. As a result, the DT model performs better.

§

Fig. 3 summarizes the result of Lemma 1 with a decision tree that can be used to determine when to use the DT or the LQE model.

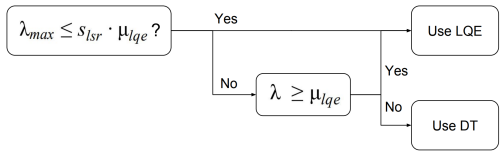

Fig. 3. Lemma 1 expressed as a decision tree.

We illustrate the practical application of Lemma 1 to determine the right design using a real HPC application. Suppose that our system uses the NIC Solarflare Flareon Ultra SFN7122F. This NIC provides hardware rings that can hold 104 buffers with each buffer consisting of 65,536 bytes worth of packets. Assuming an architecture with 20 application threads, this leads to a total buffer size of $s_{lsr} = 136,314,880$ bytes. Table 1 presents the maximum time one application thread can take ($s_{lsr}/\lambda_{max}$) to process a packet without dropping any packet in the LSR ring for a variety of burst rates ($\lambda_{max}$) from 1 Gbps to 10 Gbps.

Table 1. Maximum packet processing time for a Solarflare SFN7122F NIC

| $\lambda_{max}$ (Gbps) | 1 | 2 | 4 | 6 | 8 | 10 |
|---|---|---|---|---|---|---|
| $s_{lsr}/\lambda_{max}$ (secs) | 1.09 | 0.55 | 0.27 | 0.18 | 0.14 | 0.11 |

Table 2 and Fig. 4 provide the distribution of the packet processing times incurred for the case where the application thread runs the Bro network analyzer. These measurements were performed using a traffic dataset generated from a mix of real packet traces from our corporate network in New York and synthetically generated traffic using an Ixia traffic generator, resulting in a dataset of human generated traffic (for applications such as HTTP/HTTPS) and machine generated connections (for services such as SNMP). Table 3 summarizes the traffic dataset main statistics.

Table 2. *Packet processing time distribution.*

| [0, 10us) | [10us, 100us) | [100us, 1ms) | [1ms, 10ms) | [10ms, 100ms) |
|---|---|---|---|---|
| 305 | 405493 | 3387846 | 127 | 7 |
| **Total packets:** 3793778 | | | | |

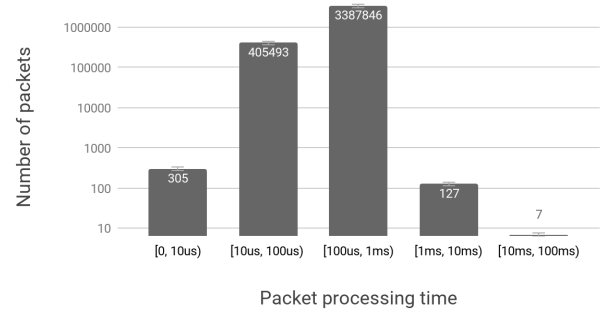

Fig. 4. Bro's packet processing time distribution (in log scale).

According to this sample, all packets can be processed by Bro in less than 100 milliseconds (the vast majority of the packets can be processed in less than 1 millisecond), that is, $1/\mu_{lqe} < 0.1$ seconds. Since the SF card yields a value of $s_{lsr}/\lambda_{max} = 0.11$ seconds at 10 Gbps, we have that $s_{lsr}/\lambda_{max} \geq 1/\mu_{lqe}$ and hence we can conclude that the long queue emulation model is the right design for the Bro network analyzer when using the Solarflare HPC NIC.

Table 3. *Statistics of the traffic dataset.*

| TCP | UDP | ICMP | Other | Avg pkt size (B) | Avg conn size (KB) |
|---|---|---|---|---|---|
| 92.34% | 7.5% | 0.02% | 0.04 | 510.25 | 7050.16 |

One final parameter to determine is the size of the software queue (USQ). While effectively this queue can be arbitrarily large (its hardware limit is given by the memory in the host), experiments demonstrate the existence of an optimal size value. This is illustrated in Fig. 5, where we feed our LQE implementation with the traffic set described in Table 3 at 10 Gbps. The results illustrate that packet drops are minimized when the number of buffers in the queue is equal to 1500. Since our implementation is based on the Solarflare NIC, where each buffer has 65,536 bytes, this corresponds to an optimal queue size of 98,304MB.

To understand the existence of this optimal queue size value, consider the two extreme cases. Assume first that the size of the queue is very small, clearly this is suboptimal as the queue will not be able to accommodate for packet bursts. Assume instead that the size of the queue is infinitely large. In this case, the queue can store a very large packet burst, but in the limit, doing so will have a negative impact on the local cache because not all the packet descriptors stored in the queue will fit in it. That is, when we try to absorb an arbitrarily large number of packets, the system's productivity is deteriorated due to an increase in cache thrashing. It's worth noticing that this optimal value depends primarily on static system parameters such as the size of the (hardware) local cache or the size of the (software) packet descriptors. Hence, this value should be fairly stable across different types of input traffic. This optimum however will differ if the hardware architecture changes (e.g., an increase in the cache size will generally imply an increase in optimal queue size) or if the packet descriptor data structure changes.

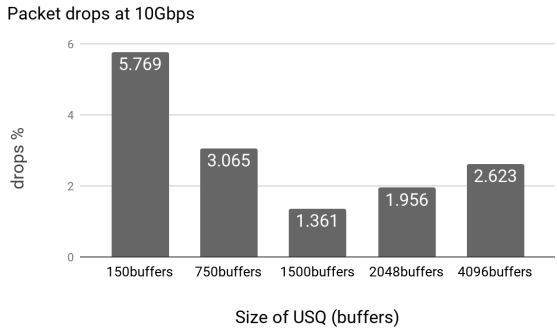

Fig. 5. Identifying the optimal size of the software queue in the LQE model.

### B. Lockless Bimodal Queues for Selective Packet Capture

We now turn our focus to address a different issue in the design of a high performant packet path. In addition to the processing of packets by the application threads, suppose that we need to support storing the packets received into disk. Because at high speed rates this can be an overwhelming task (e.g., 10 Gbps of traffic throughput can lead to the processing of up to 20 million packets per second), we will limit our specifications to capture only a finite batch of packets. We will use the name *selective packet capture* (SPC) to refer to the function of performing this type of packet capturing at very high speed rates. The SPC engine we aim at implementing should work as follows:

- After the application thread completes processing a packet, it stores the packet in a second ring. If the ring is full, its oldest packet is removed to make space for the new packet.
- The application thread has the capability to trigger a *packet capture* operation at any time. For instance, the application can decide upon processing a packet that a cybersecurity attack is being carried out and trigger the capture operation in order to save a batch of packets in the disk, allowing for a more detailed offline analysis of the suspicious packets.
- Upon triggering a packet capture operation, an SPC thread (ST) wakes up, transfers a given amount of packets from the ring to the disk and goes back to sleep.

The SPC workflow is illustrated in Fig. 6 as an extension to the LQE model described in Section II.A.

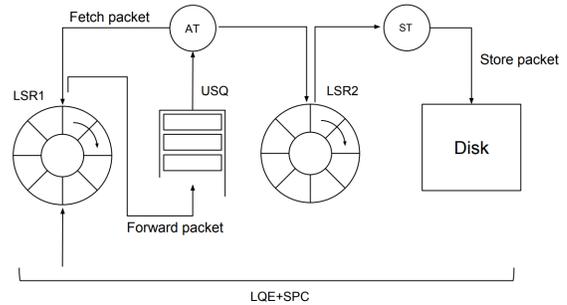

Fig. 6. Extending the LQE model to support selective packet capture (SPC).

The subsystem formed by the application thread (AT), the LSR2 ring and the SPC thread (ST) define a traditional consumer-producer problem with one caveat: the consumer is not always active. This implies that the ring needs to support two different modes of operation, one in which the consumer is sleeping without pulling any packets from the ring and another one in which the consumer is actively pulling packets from the ring. Our objective here is to design a high-performance queue supporting these two modes of operation without using locks that would negatively affect the performance of the data path. We will call such data structure a *lockless bimodal queue* (LBQ), as illustrated in Fig. 7.

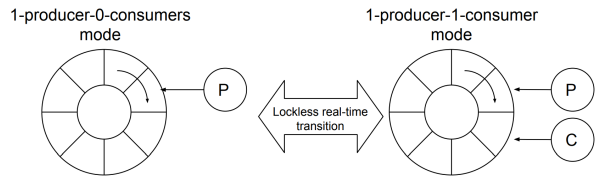

Fig. 7. A lockless bimodal queue.

We start by considering first the standard consumer-producer problem. This well-known case can be efficiently resolved without locking the ring:

---

**Lockless 1-producer-1-consumer queue**

```
1  typedef struct {
2    volatile unsigned int offset_p;
3    volatile unsigned int offset_c;
4    packet_t* vector[RINGSIZE];
5  } ring_t;
6
7  void enqueue(ring_t* ring, packet_t* pkt) {
8    while(ring->offset_p == ring->offset_c);
9    ring->vector[ring->offset_p++] = pkt;
```

```
10 }
11
12 packet_t* dequeue(ring_t* ring) {
13   if(ring->offset_p == ring->offset_c)
14     return NULL;
15   ring->offset_c = ring->offset_c + 1 % RINGSIZE;
16   return(vector[ring->offset_c - 1]);
17 }
```

In our case, the application thread acting as the producer is responsible for calling `enqueue()` whereas the SPC thread acting as the consumer calls `dequeue()`. The above solution, however, assumes the consumer is permanently active, since otherwise the producer would stall indefinitely in line 9. In our problem, the SPC thread is by default inactive and it only becomes active when the application thread triggers a packet capture operation. Hence, our design needs to also support the case of 1 producer and 0 consumers. We can make a few adjustments to the previous code to support this case:

**Lockless 1-producer-0-consumers queue**
```
1  typedef struct {
2    unsigned int offset_p;
3    unsigned int offset_c;
4    packet_t* vector[RINGSIZE];
5  } ring_t;
6
7  void enqueue(ring_t* ring, packet_t* pkt) {
8    if(ring->offset_p == ring->offset_c)
9      dequeue(ring);
10   ring->vector[ring->offset++] = pkt;
11 }
12
13 packet_t* dequeue(ring_t* ring) {
14   if(ring->offset_p == ring->offset_c)
15     return NULL;
16   ring->offset_c = ring->offset_c + 1 % RINGSIZE;
17   return(vector[ring->offset_c - 1]);
18 }
```

The above code is essentially the same as the 1-producer-1-consumer except for lines 8 and 9 (shown in bold), which have the producer take the role of the (now sleeping) consumer in order to remove the last element from the full ring to make space for the new packet.

Notice that while both the 1-producer-1-consumer and the 1-producer-0-consumers algorithms require no locks, we still need to resolve the problem of transitioning the ring between the two modes of operation. To minimize any performance penalties, the key is to ensure such transition can happen without locking the ring. We propose two solutions to achieve this objective. The first solution requires no special hardware-aided operation but assumes the producer is permanently active in order to avoid starvation on the consumer side. The second solution is not limited by such requirement but requires using compare-and-swap (CAS), a hardware-aided operation supported by most modern processors.

The following code presents the first solution without CAS:

**Lockless bimodal queue without using CAS**
**(producer must be permanently active to avoid consumer starvation)**
```
1  typedef struct {
2    volatile unsigned int offset_p;
3    volatile unsigned int offset_c;
4    volatile bool req; // owned by consumer
5    volatile bool ack; // owned by producer
6    packet_t* vector[RINGSIZE];
7  } ring_t;
8
9  void enqueue(ring_t* ring, packet_t* pkt) {
10   if(!ring->req) {
11     if(ring->ack)
12       ring->ack = false;
13     if(ring->offset_p == ring->offset_c)
14       dequeue(ring);
15   }
16   else {
17     if(!ring->ack)
18       ring->ack = true;
19     while(ring->offset_p == ring->offset_c);
20   }
21   ring->vector[ring->offset_p++] = pkt;
22 }
23
24 packet_t* dequeue(ring_t* ring) {
25   if(ring->offset_p == ring->offset_c)
26     return NULL;
27   ring->offset_c = ring->offset_c + 1 % RINGSIZE;
28   return(vector[ring->offset_c - 1]);
29 }
30
31 void start_c(ring_t* ring) {
32   ring->req = true;
33   while(!ring->ack);
34 }
35
36 void stop_c(ring_t* ring) {
37   ring->req = false;
38   while(ring->ack);
39 }
```

The main idea behind the above code is the introduction of two new functions, `start_c()` and `stop_c()`, which are to be invoked by the consumer right after it wakes up and right before it goes back to sleep, respectively. Using a two-way handshake implemented with the flags `req` and `ack`, the consumer and the producer synchronize the transition from one operational mode to another without the need for locks. Notice that to complete a transition, this approach requires the producer to be actively putting packets to the ring since the functions `start_c()` and `stop_c()` invoked by the consumer will not complete unless the producer calls `enqueue()` and executes lines 18 and 12, respectively.

In the context of high performance computing, the above implementation is often appropriate because many applications operate with producers that are constantly enqueuing packets to the ring, hence satisfying the assumption that the consumer needs to be permanently active. For applications violating this assumption, we can replace the two-way handshake operation with a compare-and-swap instruction to control the transition from one operational mode to another. This approach is presented next:

**Lockless bimodal queue using CAS**
**(producer does not need to be permanently active)**
```
1  typedef struct {
2    volatile unsigned int offset_p;
3    volatile unsigned int offset_c;
4    volatile bool trans; // used to transition modes
5    volatile bool state; // the current mode
6    packet_t* vector[RINGSIZE];
7  } ring_t;
8
9  void enqueue(ring_t* ring, packet_t* pkt) {
```

```
10    while(!cas(&ring->lock, false, true));
11    if(!ring->state) {
12      if(ring->offset_p == ring->offset_c)
13        dequeue(ring);
14      else
15        while(ring->offset_p == ring->offset_c);
16    ring->trans = false;
17    ring->vector[ring->offset_p++] = pkt;
18  }
19  packet_t* dequeue(ring_t* ring) {
20    if(ring->offset_p == ring->offset_c) return NULL;
21    ring->offset_c = ring->offset_c + 1 % RINGSIZE;
22    return(ring->offset_c - 1);
23  }
24
25  void start_c(ring_t* ring) {
26    while(!cas(&ring->trans, false, true));
27    ring->state = true;
28    ring->trans = false;
29  }
30
31  void stop_c(ring_t* ring) {
32    while(!cas(&ring->trans, false, true));
33    ring->state = false;
34    ring->trans = false;
35  }
```

In the above code, because the CAS operation is atomically executed, the consumer can change the operational mode of the ring without the need to negotiate the transition with the producer.

Because compare-and-swap is an operation widely supported by modern processors, our choice is to use the LBQ algorithm with CAS.

*C. Tail Early Dropping*

When performing network analysis, not all packets are equally important. A common example is encrypted traffic, which in general cannot be processed by the application threads since it can't be unencrypted. A source of inefficiency in today's network analysis stacks comes from the fact that by the time the application thread realizes a packet cannot be processed, such packet has already consumed important system resources. For instance, in the LQE model (Fig. 2), packets need to be moved from the LSR ring to the USQ queue, and then picked up by the application thread before they can be processed. If the architecture includes the selective packet capture module (Fig. 6), then the packet also needs to be moved to the LSR2 queue. Each of these steps consume both computational and memory resources that yield no benefit if the packet needs to be ultimately dropped.

A general principle in the design of high performance network analyzers is that if a packet is to be discarded from the analysis, then it should be dropped as early as possible. To enable this principle, we develop a module called *tail early dropping* (TED). TED is a queuing policy that allows for entire connection tails to be dropped once the analyzer threads conclude that such connections are no longer relevant to the analysis. This technique allows also to prioritize the front of a connection (which typically includes more relevant information) when the system is congested, by dropping connection tails. Next, we describe this optimization in a bit more detail.

Fig. 8 provides a diagram of our LQE model extended with the TED component. TED is composed of a connection cache and a decision module implementing the following algorithm:

```
TED
1   Upon receiving a packet, do:
2     conn = lookup_connection_table(packet)
3     if conn.shunt or conn.packet_rec > ted_thr:
4       drop the packet
5     else:
6       forward the packet
7   Periodically, do:
8     if system is congested:
9       ted_thr = min(ted_thr / 2, ted_min);
10    else:
11      ted_thr += 1;
```

The algorithm is composed of two parts: a packet forwarding routine that runs every time a packet is received (lines 1 through 6) and a housekeeping routine that runs periodically (lines 7 through 11). The packet forwarding routine decides to forward a packet only if the connection (`conn`) associated with this packet has not been marked for shunting (`conn.shunt`) and the total number of packets received from this connection (`conn.packet_rec`) does not exceed a threshold (`ted_thr`). If one of these two conditions is not met, the packet is dropped.

The housekeeping routine maintains a TED threshold parameter `ted_thr` as follows: if the system is congested, then the value of `ted_thr` is reduced by half down to a minimum value of `ted_min`; otherwise, the threshold is increased by one unit. This threshold provides a mechanism to dynamically cut connections tails more aggressively if the system is congested, effectively giving higher priority to the front of the connections.

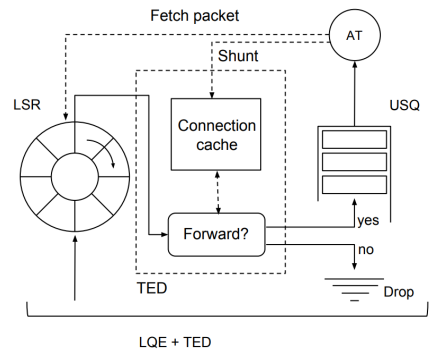

Fig. 8. Extending the LQE model with tail early dropping (TED).

Hence, the TED module enables two mechanisms to reduce the volume of ingested traffic by the application thread: through a *shunting* mechanism triggered by the application thread itself when it detects that a connection no longer needs to be processed (for instance, this covers the case of encrypted connections) or through a dynamic mechanism that prioritizes the packets at the front of a connection depending on the level of system congestion. In our implementation, we also use packet drops at the LSR ring to determine if the system is congested. Hence, we implement line 8 in the algorithm by using this condition:

```
8      If the LSR ring is dropping packets:
```

Fig. 9 illustrates the benefits of using TED queuing through a benchmark. In this case we test the performance o a single Bro worker (one application thread) with and without the TED queuing optimization. Using httperf [4], we synthetically create a packet trace consisting of a population of HTTP clients downloading a 1MB file from 25 different HTTP servers. With this setup, we collect a 65 GB trace which we use to stress our implementation by replaying it at various rates. Fig. 9 presents the amount of HTTP and file events generated by the Bro worker when processing the packet trace at 500 Mbps and 5 Gbps. In all our tests, events are measured in terms of the number of log records generated by Bro. As shown, while at 500Mbps both configurations (with and without TED) are capable of extracting practically 100% of all the events, at 5 Gbps the TED configuration delivers 2.5X more HTTP events and 3X more file events.

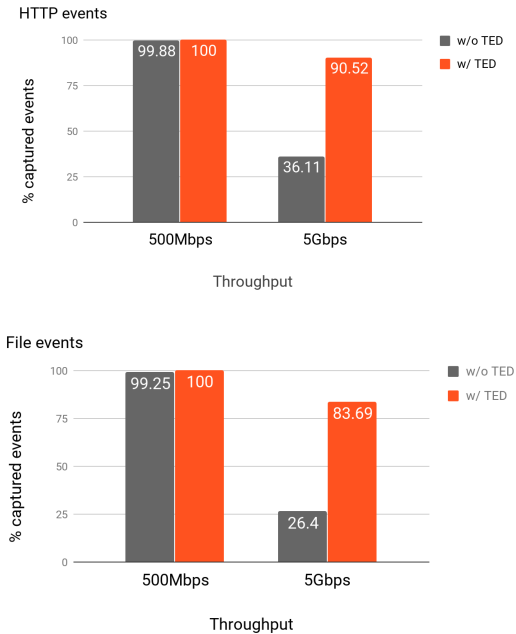

Fig. 9. TED queuing performance.

### D. LFN Tables

Next, consider the problem of analyzing and optimizing the performance of the connection cache in the TED module (Fig. 8). In its most general form, this is a data structure shared by all application threads which provide feedback down to the lower layer indicating whether packets of a given connection should be forwarded or shunted. As multiple threads can write to this cache, its access needs some form of synchronization to guarantee the integrity of its state. Locking this data structure while processing packets at very high speed rates is prohibitive since that would stall application threads as they try to gain access to the cache, which would put back pressure down to the LSR ring and lead to packet drops. Once again we need to design a suitable data structure that allows for parallel write access to the table without requiring a lock.

Toward this objective, we designed a data structure called *lock-free low-false negative* (LFN) table. The LFN data structure defines a family of hash tables and sets that can achieve lockless concurrent access to shared state by trading off a low probability of false negatives and a very low (or negligible) probability of false positives. These data structures were formally introduced in [5] and we herein provide a summarized description and explain how to apply them to implement the connection cache.

In its most basic form, an LFN data structure implements the following `put` and `get` methods to store and retrieve a key `k` associated with a value `v`:

```
Initial state:  T[e] = NULL for all e such that 0 ≤ e < n;
Parameters:
  n: size of the table
  l: processor's integer space size (typically 2^32 or 2^64)
h(x, k): the hash value of k modulo x
cat(x, y): concatenates the bytes from and x and y
put(k, v)
1   T[h(n, k)].value = v
2   T[h(n, k)].hash = h[l, cat(k, v)]
get(k)
3   if T[h(n, k)].hash == h(l, cat(k, T[h(n, k)].value)):
4       return T[h(n, k)].value
5   else:
6       return NULL
```

The basic idea of an LFN hash table is that it can avoid using locks by leveraging the processor's capability to perform integer operations atomically. Specifically, line 2 in the `put` method is guaranteed to be executed atomically, which ensures that the value of `T[h(n, k)].hash` will be coherent. For instance, on a 64 bit processor, `T[h(n, k)].hash` can be represented using a 64 bit integer and so $l = 2^{64}$.

LFN hash tables can still have collisions, but they are designed in a way that when they happen, with high probability the `get` operation will return a `NULL` if the key we seek to find has been evicted by another key. More specifically, while a `put` operation is always successful, a `get` can lead to three states:

- The key `k` is stored in the table and it's value is correctly returned by the `get`, or the key is not stored in the table and the `get` operation returns `NULL`. This is called a true state.
- The key `k` is stored in the table but it is no longer found and the `get` returns `NULL`. This is called a false negative state.
- The `get` operation returns the value of another key `k'` different than `k`. This is called a false positive state.

Clearly the desired outcome is a true state. While this cannot be guaranteed at all times, an LFN table ensures that false negatives happen with low probability and false positives with extremely low probability. Mathematically, if there are *k* keys stored in the table, then a false negative occurs with a probability $(k − 1)/2n$ while a false positive occurs with a probability $(k − 1)/2l$ (see [5]). For instance, storing one thousand keys ($k = 1000$) in a table of 1 million entries ($n = 10^6$) using a 64 bit integer architecture ($l = 2^{64}$), we have that the chances of an element in the table to be in a false negative or false positive state are $5 \cdot 10^{-4}$ and $2.7 \cdot 10^{-17}$, respectively. Fig. 10 provides a graphical representation of this result for different values of *k, n*, and *l*.

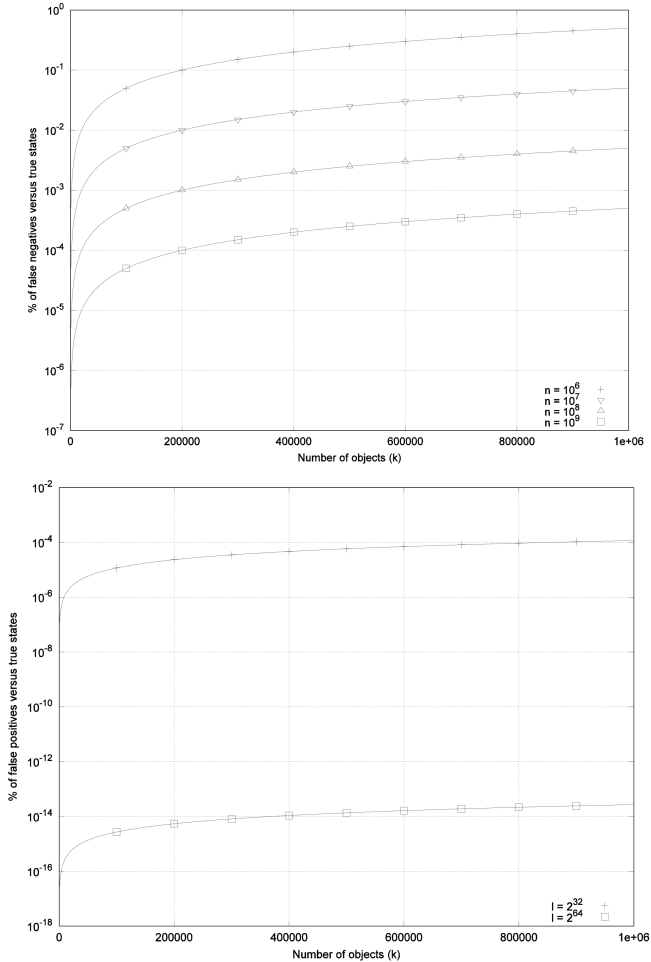

Fig. 10. False negative (top) and a false positive (bottom) probability of an LFN table.

LFN tables have interesting properties when used in the context of per-flow state tracking problems commonly found in computer networks. First, as we use them to keep the shunting state of all the active connections, it's critical that the system does not drop connections that should otherwise be analyzed. This is especially important in the case of cyber security analysis. Luckily, this case corresponds to the probability of false positive, which as we showed it's extremely low. While false negatives can occur with higher probability, they correspond to the case that a connection is not shunted while it should be shunted. Hence, LFN tables can be understood as data structures with asymmetric performance erring on the safe side.

Since their size is fixed, another good property of LFN tables is their resilience to denial of service (DoS) attacks. Notice that this is not the case with other hashing table schemes such as *separate chaining,* which grow with the number of input connections and lead to state explosion upon a DoS attack.

A final interesting property of LFN tables is that they eliminate the need for inactivity timeouts that are typically necessary to manage connection tables. For instance, stateless protocols like UDP do not signal the end of a connection, and so one needs to rely on inactivity timeouts to clean them up from the table. LFN tables however perform the cleanup operation in a natural way, as every stale connection will eventually be replaced by a new one through a key collision. This also implies that, when dealing with network connections, the real false negative rate will be smaller than $(k-1)/2n$, because connections have a limited lifetime. In particular, a real false negative will only occur if a new incoming connection collides with an existing connection *that is still active*. If the existing connection is already terminated, then the collision does not yield a false negative, instead it performs a natural clean up operation by removing the old connection from the table and replacing it with the new connection. When implementing an LFN table, it's important to choose its size parameter *n* large enough such that the duration of the connections is taken into account, ensuring the table operates in a state where key collisions correspond to natural connection cleanups instead of false negatives.

*E. Multiresolution Priority Queues to Manage Timers*

We now turn to describing another data structure designed to help eliminate a different system wide bottleneck. Many real time network analyzers require the implementation of timers to keep track of state. For instance, every time a TCP or a UDP connection is processed, the Bro network analyzer allocates a few timers. Examples include the connection establishment, the connection inactivity or the connection linger timeouts, among several others. At rates of 10 Gbps, the system needs to process tens of thousands of connections per second, requiring to manage in the order of hundreds of thousands of timers per second. The management of timers is commonly carried out using a priority queue where the expiration time of each timer is treated as its priority in the queue. (See for example Section 6.5 of [6].) In this way the first element of the queue corresponds to the timer that is to expire next among all the timers in the queue. Traditionally, the priority queue is implemented using a binary heap, which has a computational cost to insert and remove elements of $O(log(n))$, where *n* is the number of elements in the heap.

While binary heaps are excellent implementations of priority queues, we find that when dealing with very high speed traffic, they still become a system bottleneck. This is illustrated in the next list of Bro functions ordered by their computational cost when running it against the traffic dataset introduced in Table 3 at 10 Gbps and as measured by *gperftools*, the CPU profiler developed by Google [7]:

```
Total: 63724 samples
    4139    6.5% PriorityQueue::BubbleDown
    2500    3.9% SLL_Pop
    1899    3.0% Ref
    1829    2.9% Unref
    1701    2.7% PackedCache::KeyMatch
    1537    2.4% Attributes::FindAttr
    1249    2.0% Dictionary::Lookup
    1184    1.9% NameExpr::Eval
```

As shown, the function `PriorityQueue::BubbleDown` takes the top spot with a cost of 6.5% of the total processing cost. This function implements the standard *bubble down* operation (also know in the literature as the *heapify* procedure [6]) part of the binary heap based implementation of a priority queue. Specifically, Bro uses this method every time it needs to insert or remove an element from the queue of timers. To overcome this bottleneck, we developed *multiresolution priority queues* (MRPQ), a data structure that achieves greater performance than the standard binary heap based implementation by trading off a controllable amount of resolution in the space of priorities.

Introduced in more detail in [8], a multiresolution priority queue breaks the information-theoretic barriers of the problem of ordering $n$ elements according to their priorities by exploiting the multiresolution properties of the priority space. Since in many problems the entropy of the priority space is lower than the entropy of the key space, the end result is a container data structure with a lower computational complexity. In particular, if the space of priorities is multi-resolutive, its entropy will be independent of the number of elements in the queue, and hence the ordering problem can be resolved in constant time $O(1)$. (See [8] for a detailed proof.) This makes the resulting data structure substantially more efficient than a binary heap.

Since all Bro timers have an expiration value of 1 second or higher, its priority queue operates on a multi-resolutive priority space of resolution 1 second. Hence, we can use a multiresolution priority queue to improve the performance of Bro's timer manager without losing accuracy. Table 4 summarizes the computational costs savings in the processing of timers due to using a multiresolution priority queue instead of a binary heap.

Table 4. *Computational cost reductions in the processing of timers.*

| Algorithm | Insert | Peek | ExtractMin | Extract |
|---|---|---|---|---|
| Binary heap | $log(n)$ | $O(1)$ | $log(n)$ | $log(n)$ |
| Multiresolution priority queue | $O(1)$ | $O(1)$ | $O(1)$ | $O(1)$ |

We extended Bro with a multiresolution priority queue configured to support timers with 1 second resolution. Fig. 11 presents the results of benchmarking Bro against the traffic data set in Table 3 at a variety of speeds. At a microprocessor level, Fig. 11-top shows how the MRPQ data structure achieves better cache performance by reducing system wide cache miss ratios from 21% to 17% at 10 Gbps rates. This is due mainly to the function `PriorityQueue::BubbleDown` which requires scanning through $O(log(n))$ timers every time the timer queue is accessed (for both insert and remove operations). This leads to an increase in the amount of cache thrashing and cache misses. Instead, the MRPQ data structure requires $O(1)$ operations to access the queue, hence only touching the element that is to be inserted or removed, resulting into a better cache performance. As shown in Fig. 11-middle and Fig. 11-bottom, at a system level this leads to a reduction in packet drops and an increase in the number of events generated by the network analyzer. As in Section II.C, here we also measure events as the number of log records generated by Bro.

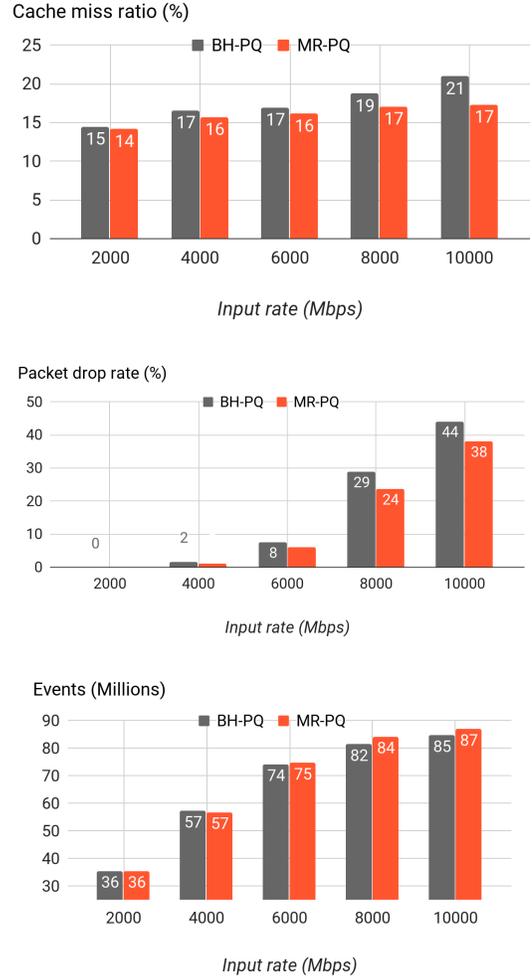

Fig. 11. Performance improvements achieved in Bro when using a multiresolution priority queue to manage timers.

### III. BENCHMARKS

We have implemented all the optimizations described in this paper as part of R-Scope, Reservoir Lab's high speed network appliance that runs a Bro engine at its core. Fig. 12 presents benchmarks measuring the benefits of these optimizations. All tests were performed using the traffic data set described in Table 3 by replaying each test for a duration of 10 minutes at speeds ranging from 2 Gbps through 10 Gbps. In all configurations, a Bro cluster with 20 application threads (known as Bro workers) was employed. Results for three different configurations are presented:

- *Stock Bro myri* corresponds to the standard Bro distribution without any of the optimizations in this paper and using the Myricom 10G-PCIE2 NIC. This configuration uses Myricom's optimized libpcap library to deliver packets from the wire to the application threads.
- *R-Scope Myri/mCore* corresponds to the standard Bro distribution with all the optimizations in this paper except for multiresolution priority queues (MRPQ) and long queue emulation (LQE). The optimizations are referred with the codename *mCore*. This configuration also uses Myricom's libpcap library (hence, LQE is not being used as mentioned.)

- *R-Scope SF/mCore* runs the standard Bro distribution with all the optimizations on this paper and using the Solarflare Flareon Ultra SFN7122F NIC. This configuration uses Solarflare's native API instead of libpcap. The optimizations are referred with the codename mCore+.

While the hardware configuration is identical for the three configurations, the R-Scope Myri/mCore and R-Scope SF/mCore configurations include also additional optimizations implemented as part of the R-Scope appliance that are not described in this paper. For instance, the R-Scope SF/mCore configuration directly runs on the Solarflare native API, bypassing the libpcap library. So while the optimizations described in this paper account for a substantial fraction of the performance gains shown in Fig. 12, these aggregated benchmarks should be qualitatively interpreted.

Results in Fig. 12 are shown in terms of the number of total events (measured by counting the total number of log records as we did in Sections II.C and II.F) and total number of reported connection records. At 10 Gbps, speed ups of 5.1X and 7.8X are achieved for the number of events and connection records, respectively. Similar improvements are achieved across all types of metadata reported by Bro.

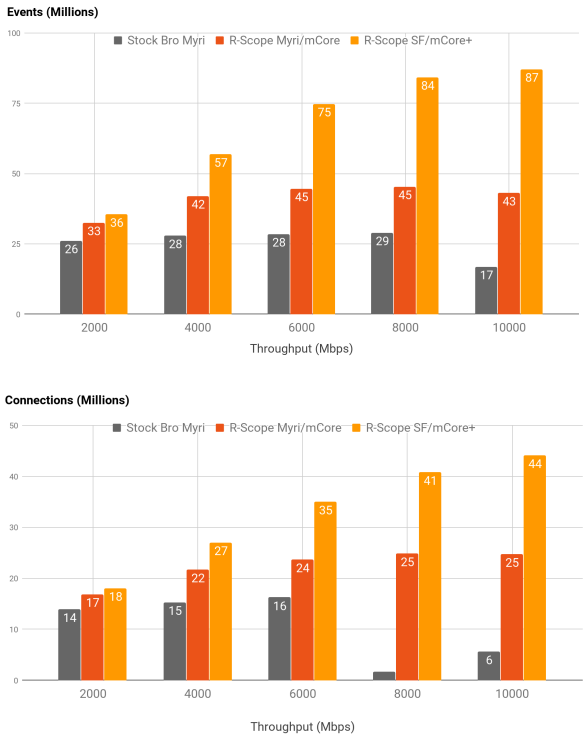

Fig. 12. Benchmarks corresponding to the three different configurations.

## IV. CONCLUSIONS

Table 5 provides a summary of each the optimizations presented in this paper and their main benefit. These algorithms and data structures have been developed as a multi-step optimization process spanning multiple years of research and development. At each step, we considered the various potential bottlenecks by using a variety of methods including (1) performing fine-grained as well as coarser aggregated benchmarks, (2) measuring code performance through a CPU profiler, (3) taking measurements from hardware performance counters or (4) directly adding measurement code, among other techniques. At each step a new bottleneck is unveiled, and then a new algorithm or data structure is designed to eliminate it while ensuring a net positive performance gain at a system wide level.

Table 5. Summary of algorithms and data structures

| Algorithm/data structure | Benefit |
|---|---|
| Long queue emulation | Reduces packet drops from fixed-size hardware rings |
| Lockless bimodal queues | Improves packet capturing performance |
| Tail early dropping | Increases information entropy and extracted metadata |
| LFN tables | Reduces state sharing overhead |
| Multiresolution priority queues | Reduces cost of processing timers |

The solutions we present are generally applicable to problems outside the area of network analysis. For instance, priority queues are key data structures core to many HPC applications in the field of computer science, including graph theory problems such as the shortest path, Huffman compression codes, operating systems, Bayesian spam filtering, discrete optimization, simulation of colliding particles, or artificial intelligence to name some examples. With a lower computational cost, the proposed multiresolution priority queue can be used to address these problems if they define a multi-resolutive priority space or if they are tolerant to small errors.

LFN tables can also be generally applied to the problem of efficiently tracking per-flow state at high speed rates, which is also commonly found in other network equipment such as routers and firewalls.

Finally, LQE, LBQ and TED are algorithms that can be generally applied to the problem of efficiently moving packets between various queues, allowing to tune the data path towards identifying sweet spots in the continuum defined by the trade-off performance versus accuracy.

REFERENCES

[1] V. Paxson, "Bro: a system for detecting network intruders in real-time," *Computer Networks*, vol. 31, no. 23–24, pp. 2435–2463, 1999.
[2] J. Ros-Giralt, A. Commike, D. Honey, and R. Lethin, "High-performance many-core networking," in *Proceedings of the Second Workshop on Innovating the Network for Data-Intensive Science - INDIS '15*, 2015.
[3] A. Leon-Garcia, *Probability, Statistics, and Random Processes For Electrical Engineering*. Pearson Higher Ed, 2011.
[4] D. Mosberger, T. Jin, and H.-P. Laboratories, *Httperf: A Tool for Measuring Web Server Performance*. 1998.
[5] J. Ros-Giralt, A. Commike, R. Rotsted, P. Clancy, A. Johnson, and R. Lethin, "Lockless hash tables with low false negatives," in *2014 IEEE High Performance Extreme Computing Conference (HPEC)*, 2014.
[6] T. H. Cormen, *Introduction to Algorithms*. MIT Press, 2009.
[7] "Google Performance Tools." [Online]. Available: https://github.com/gperftools/gperftools. [Accessed: 17-Sep-2017].
[8] Jordi Ros-Giralt, Alan Commike, Peter Cullen, Jeff Lucovsky, Dilip Madathil, Richard Lethin, "Multiresolution Priority Queues," presented at the IEEE High Performance Extreme Computing Conference, Boston, USA.